\documentclass[prl,twocolumn,floatfix,amssymb,showpacs]{revtex4}

\usepackage{graphicx}
\usepackage{amsmath}

\newcommand{\bB}{\boldsymbol{\mathcal{B}}}
\newcommand{\mcB}{\mathcal{B}}

\begin{document}

\title{Group theoretical analysis of double acceptors in a
magnetic field: identification of the {S}i:{B}$^+$ ground state}

\author{G.~D.~J.~Smit}
\email{g.d.j.smit@tnw.tudelft.nl}
\author{S.~Rogge}
\email{s.rogge@tnw.tudelft.nl}
\author{J.~Caro}
\author{T.~M.~Klapwijk}
\affiliation{Department of NanoScience, Delft University of
Technology, Lorentzweg 1, 2628 CJ Delft, The Netherlands}

\begin{abstract}
A boron impurity in silicon binding an extra hole is known to have
only one bound state at an energy of just below 2~meV. The nature
of the Si:B$^+$ ground state is however not well established. We
qualitatively analyze the behavior in a magnetic field of isolated
acceptors in a tetrahedral lattice binding two holes using group
theory. Applying these results, we analyze recent measurements and
conclude that the ground state of B$^+$ is most compatible with a
non-degenerate $\Gamma_1$ state.
\end{abstract}

\date{\today}


\maketitle

\section{Introduction}

A neutral boron acceptor in silicon is able to weakly bind an
extra hole, resulting in a positively charged ion (B$^+$). This
entity is an example of a positively charged acceptor, commonly
denoted by A$^+$, which is the counter part of the better known
negatively charged donor D$^-$. Both are related to their
atomic-physics analogue, the negative hydrogen ion H$^-$. The
energy states associated with these ions are very shallow and
spatially large. When their concentration is sufficiently high,
their overlapping wave functions can form an upper Hubbard band
\cite{norton76} and they play an important role in electronic
transport in semiconductors at low temperatures. Since a few
years, electronic states of individual dopant atoms gained renewed
interest due to their prospective application in Si-based solid
state quantum computing \cite{kane98}.

Neither theoretically nor experimentally much work has been done
on the B$^+$-state. Optical spectroscopy is difficult due to the
small ionization energy (less than 2~meV \cite{burger84}). In
particular, the nature of its ground state is not well-established
and to our knowledge no results have been published on the
magnetic field dependence of the energy levels.

The purpose of this paper is twofold. First, we present a general
group-theoretical study of the magnetic field dependence of
two-hole states in tetrahedral semiconductors. To our knowledge,
such an analysis has not been published before. Second, because
our analysis includes all possibilities for the B$^+$ ground
state, it enables us to compare our results with our recently
published measurements of B$^+$ in a magnetic field \cite{caro03}
and to draw conclusions about the nature of the B$^+$ ground
state.

\section{Background}

The nature of the energy levels of a \emph{neutral} boron acceptor
(B$^0$) in silicon is well-known \cite{ramdas81} and the Zeeman
effect in B$^0$ has been studied in detail, both theoretically
\cite{bhattacharjee72} and experimentally \cite{merlet75}. The
B-impurity is located at substitutional sites of the tetrahedral
silicon lattice. The (one-hole) ground-state is a 1$s$-like
fourfold degenerate state that belongs to the $\Gamma_8$
representation of the tetrahedral double group $\bar{T}_d$ (for
the nomenclature of representations used, see
Table~\ref{tab:charTd}). The bound hole has total angular momentum
$j=\frac{3}{2}$. The single-hole wave function is the product of a
1$s$ hydrogen-like envelope function and a band-like function. Due
to spin-orbit interaction in silicon, the $j=\frac{1}{2}$ valence
band is split off by $\sim 43$~meV \cite{ramdas81} and does not
need to be considered in first order. A magnetic field completely
lifts the fourfold degeneracy and the lowest order Zeeman effect
of the $\Gamma_8$ state is linear.

\begin{table}
  \caption{Character table for the double group $\bar{T}_d$.}
  \label{tab:charTd}
  \centering
  \begin{tabular}{l|cccccccc}
    \hline
    \hline
       & $E$ & $\bar{E}$ & $8C_3$ & $8\bar{C}_3$ & $3C_2$,
       $3\bar{C}_2$ & $6S_4$ & $6\bar{S}_4$ & $6\sigma_d$,
       $6\bar{\sigma}_d$ \\
    \hline
    $\Gamma_1$ & 1 & 1 & 1 & 1 & 1 & 1 & 1 & 1 \\
    $\Gamma_2$ & 1 & 1 & 1 & 1 & 1 & $-1$ & $-1$ & $-1$ \\
    $\Gamma_3$ & 2 & 2 & $-1$ & $-1$ & 2 & 0 & 0 & 0 \\
    $\Gamma_4$ & 3 & 3 & 0 & 0 & $-1$ & 1 & 1 & $-1$ \\
    $\Gamma_5$ & 3 & 3 & 0 & 0 & $-1$ & $-1$ & $-1$ & 1 \\
    \hline
    $\Gamma_6$ & 2 & $-2$ & 1 & $-1$ & 0 & $-\sqrt{2}$ & $\sqrt{2}$ & 0 \\
    $\Gamma_7$ & 2 & $-2$ & 1 & $-1$ & 0 & $\sqrt{2}$ & $-\sqrt{2}$ & 0 \\
    $\Gamma_8$ & 4 & $-4$ & $-1$ & 1 & 0 & 0 & 0 & 0 \\
    \hline
    \hline
  \end{tabular}
\end{table}

As far as symmetry is concerned, the B$^+$-state is similar to
neutral group-II acceptors in a tetrahedral lattice, which are
well-studied (e.g. Ref.~\onlinecite{kartheuser73} and references
therein). Coupling two $j=\frac{3}{2}$ ($\Gamma_8$) holes bound to
a single nucleus gives rise to a six-fold degenerate state,
because due to the Pauli-principle only the \emph{anti}-symmetric
part of $\Gamma_8\times\Gamma_8$ must be taken into account. This
can be reduced to its components as $\{\Gamma_8\times\Gamma_8\} =
\Gamma_1+\Gamma_3+\Gamma_5$. Interaction between the two holes can
split the state into a non-degenerate $\Gamma_1$ state with total
angular momentum $J=0$ and a fivefold $\Gamma_3+\Gamma_5$ state
carrying $J=2$.

Detailed quantitative calculations, which are necessary to
establish the ordering and splitting of these levels, are very
difficult to carry out, because of the many complicated physical
effects that must be taken into account (valence band structure,
crystal field, Jahn-Teller-effect, etc.). Hund's rule, well-known
from atomic physics, predicts that the more symmetric $\Gamma_1$
state has a higher energy than the $\Gamma_3+\Gamma_5$ state, such
that the latter is the ground state. The same conclusion was drawn
from numerical calculations based on effective mass theory
\cite{rodina93}. However, it has been shown that a dynamic
Jahn-Teller effect can provide a mechanism to reverse the ordering
of the levels \cite{ham93,ham95}, leading to a $\Gamma_1$ ground
state. This has in fact been observed in several neutral double
acceptors.

Very little experimental work on B$^+$ has been done. The binding
energy of the second hole in an isolated B$^+$-state has been
measured in phonon-induced conductivity (PIC) measurements
\cite{burger84} and photoconductivity experiments
\cite{sugimoto79}. It is slightly below 2~meV. Stress-dependence
has been investigated with the same techniques
\cite{gross95,sugimoto79} and in one case the results were
explained as evidence for a stress-induced ground state splitting
\cite{gross95}. However, interpretation of the conductivity data
is non-trivial, because only levels which are very close to either
the ground state of B$^+$ or the valence band edge can be observed
with these techniques. Similar experiments in a magnetic field
\cite{roshko96} showed a linear increase of the binding energy,
which was ascribed to Landau level formation in the valence band.
In these experiments, no additional shift or splitting was
resolved.

Recent transport experiments in Si resonant tunneling devices
provide a way to directly observe the magnetic field dependence of
the B$^+$ state \cite{caro03}. These experiments showed a super
linear shift of the ground state towards the valence band
(Fig.~\ref{fig:Bmeas}). Neither a ground state level-splitting nor
bound excited states were observed.

\section{Double acceptors in a magnetic field}

Here, we present a group theoretical study to qualitatively
analyze the magnetic field behavior of isolated acceptors binding
two holes in a tetrahedral semiconductor for various possible
states. This analysis is not only applicable to neutral group-II
acceptors (e.g. Zn in Ge), but also to group-III acceptors binding
an extra hole and singly ionized group-I acceptors (e.g. Cu$^-$ in
Ge). After this general part, we return to the specific situation
of B$^+$.

\begin{table}
  \caption{Overview of possible two-hole states arising from
  products of two single hole states and their reduction to
  irreducible representations of $\bar{T}_d$. For states originating from
  two \emph{equivalent} single hole states (first two lines in the
  table), the Pauli principle allows only the antisymmetric part
  to be considered.}
  \label{tab:twohole}
  \centering
  \begin{tabular}{l|l}
    \hline
    \hline
    Combination & Two-hole states \\
    \hline
    $\{\Gamma_8\times\Gamma_8\}$ & $\Gamma_1+\Gamma_3+\Gamma_5$ \\
    $\{\Gamma_6\times\Gamma_6\}=\{\Gamma_7\times\Gamma_7\}$ &
      $\Gamma_1$ \\
    $\Gamma_8\times\Gamma'_8$ & $\Gamma_1+\Gamma_2+\Gamma_3+2\Gamma_4+2\Gamma_5$ \\
    $\Gamma_6\times\Gamma'_6=\Gamma_7\times\Gamma'_7$ &
      $\Gamma_1+\Gamma_4$ \\
    $\Gamma_8\times\Gamma_7=\Gamma_8\times\Gamma_6$ &
      $\Gamma_3+\Gamma_4+\Gamma_5$ \\
    $\Gamma_7\times\Gamma_6$ & $\Gamma_2+\Gamma_5$ \\
    \hline
    \hline
  \end{tabular}
\end{table}

We subsequently consider various possible two-hole levels and
analyze their behavior in a magnetic field using perturbation
theory. All such levels transform according to single-valued
representations of $T_d$, as shown in the overview in
Table~\ref{tab:twohole}. We assume that the Coulomb force and
spin-orbit interaction between the holes is sufficiently strong to
split the levels into their irreducible components. Because of its
possible importance for B$^+$, we also consider the
$\Gamma_3+\Gamma_5$ level. In all cases it is assumed that the
level under consideration is well separated from neighboring
levels.

Furthermore, we briefly address the analogue of the central field
approximation in atomic physics, where it is assumed that each of
the two holes moves in the field of the negative ionized acceptor
core and the averaged effective potential due to the other hole.
In this approximation, the symmetry of the field in which each
hole moves is unaffected by the presence of the second hole. This
method is known to give a good description for some group-II
acceptors in Si and Ge \cite{kartheuser73}.

\begin{table}
  \caption{Character table for the double group $\bar{S}_4$
  ($\omega=e^{i\pi/4}$).}
  \label{tab:charS4}
  \centering
  \begin{tabular}{l|cccccccc}
    \hline
    \hline
       & $E$ & $\bar{E}$ & $C_2$ & $\bar{C_2}$ &
         $S_4$ & $S_4^{-1}$ & $\bar{S}_4$ & $\bar{S}_4^{-1}$ \\
    \hline
    $\Gamma_1$ & 1 & 1 & 1 & 1 & 1 & 1 & 1 & 1 \\
    $\Gamma_2$ & 1 & 1 & 1 & 1 & $-1$ & $-1$ & $-1$ & $-1$ \\
    $\Gamma_3$ & 1 & 1 & $-1$ & $-1$ & $-i$ & $i$ & $-i$ & $i$ \\
    $\Gamma_4$ & 1 & 1 & $-1$ & $-1$ & $i$ & $-i$ & $i$ & $-i$ \\
    \hline
    $\Gamma_5$ & 1 & $-1$ & $-i$ & $i$ & $-\omega$ & $\omega^3$ & $\omega$ &
    $-\omega^3$ \\
    $\Gamma_6$ & 1 & $-1$ & $i$ & $-i$ & $\omega^3$ & $-\omega$ & $-\omega^3$ &
    $\omega$ \\
    $\Gamma_7$ & 1 & $-1$ & $-i$ & $i$ & $\omega$ & $-\omega^3$ & $-\omega$ &
    $\omega^3$ \\
    $\Gamma_8$ & 1 & $-1$ & $i$ & $-i$ & $-\omega^3$ & $\omega$ & $\omega^3$ &
    $-\omega$ \\
    \hline
    \hline
  \end{tabular}
\end{table}

\begin{table}
  \caption{Character table for the double groups
  $\bar{C}_3$ (left; $\omega=e^{i\pi/3}$) and $\bar{C}_{1h}$ (right).}
  \label{tab:charC3}
  \begin{minipage}[t]{5cm}
    \begin{tabular}{l|cccccc}
      \hline
      \hline
         & $E$ & $\bar{E}$ & $C_3$ & $C_3^{-1}$ & $\bar{C}_3$ & $\bar{C}_3^{-1}$ \\
      \hline
      $\Gamma_1$ & 1 & 1 & 1 & 1 & 1 & 1 \\
      $\Gamma_2$ & 1 & 1 & $-\omega$ & $\omega^2$ & $-\omega$ & $\omega^2$ \\
      $\Gamma_3$ & 1 & 1 & $\omega^2$ & $-\omega$ & $\omega^2$ & $-\omega$ \\
      \hline
      $\Gamma_4$ & 1 & $-1$ & $-\omega^2$ & $\omega$ & $\omega^2$ & $-\omega$ \\
      $\Gamma_5$ & 1 & $-1$ & $\omega$ & $-\omega^2$ & $-\omega$ & $\omega^2$ \\
      $\Gamma_6$ & 1 & $-1$ & $-1$ & $-1$ & 1 & 1 \\
      \hline
      \hline
    \end{tabular}
  \end{minipage}
  \begin{minipage}[t]{3.5cm}
    \begin{tabular}{l|cccc}
      \hline
      \hline
         & $E$ & $\bar{E}$ & $\sigma_h$ & $\bar{\sigma}_h$\\
      \hline
      $\Gamma_1$ & 1 & 1 & 1 & 1\\
      $\Gamma_2$ & 1 & 1 & $-1$ & $-1$\\
      \hline
      $\Gamma_3$ & 1 & $-1$ & $-i$ & $i$\\
      $\Gamma_4$ & 1 & $-1$ & $i$ & $-i$\\
      \hline
      \hline
    \end{tabular}

    \vfill
  \end{minipage}
\end{table}

When a magnetic field $\bB$ is applied, new terms are introduced
in the Hamiltonian of the holes, as given by the
Zeeman-Hamiltonian
\begin{equation*}
  \begin{split}
    \mathcal{H}_Z &= -\mu_B(\mathbf{L}+2\mathbf{S})\cdot\bB \\
      & -\frac{1}{2}m^*\mu_B^2\left\{(r_1^2+r_2^2)\mathcal{B}^2
      -[(\mathbf{r_1}+\mathbf{r_2})\cdot\bB]^2\right\}, \\
  \end{split}
\end{equation*}
where $m^*$ is the hole effective mass and $\mathbf{L}$ and
$\mathbf{S}$ are the total orbital and spin angular momenta in
units of $\hbar$. The quantity $\mathbf{L}+2\mathbf{S}$ is the
total static magnetic moment of the system. Moreover, $\mu_B$ is
the Bohr magneton and $\mathbf{r}_i$ is the position vector of the
$i$-th hole.

The symmetry group of the Zeeman Hamiltonian $\mathcal{H}_Z$ is
$\bar{C}_{\infty h}$. Unless $\bB$ is directed along one of the
main crystallographic axes, the symmetry group of the total
Hamiltonian $\mathcal{H}=\mathcal{H}_0+\mathcal{H}_Z$ reduces to
the trivial group. When $\bB$ is parallel to a $\langle
100\rangle$, $\langle 111\rangle$ or $\langle 110\rangle$
direction in the crystal, the symmetry group of the total
Hamiltonian reduces from $\bar{T}_d$ to $\bar{S}_4$, $\bar{C}_3$
or $\bar{C}_{1h}$, respectively. The relevant character tables are
given in Table~\ref{tab:charS4} and~\ref{tab:charC3}. Because all
resulting groups are Abelian (commutative), it follows that the
application of a magnetic field completely removes the degeneracy
of all levels \footnote{But not necessarily in a first order
approach.}. The way in which the $\Gamma_i$ levels exactly split
in a magnetic field is presented in Table~\ref{tab:Bsplit}.

\begin{table}
  \caption{Reduction of the representations of $T_d$ when a
  magnetic field is applied along a $\langle 100\rangle$,
  $\langle 111\rangle$, or $\langle 110\rangle$ direction
  of the tetrahedral lattice. From this table it can be deduced
  how the double acceptor levels split in a magnetic field.}
  \label{tab:Bsplit}
  \centering
  \begin{tabular}{l||c|c|c}
  \hline
  \hline
  Direction & $\langle 100\rangle$ & $\langle 111\rangle$ &
    $\langle 110\rangle$ \\
  Group & $S_4$ & $C_3$ & $C_{1h}$ \\
  \hline
  $\Gamma_1$ ($T_d$) & $\Gamma_1$ & $\Gamma_1$ & $\Gamma_1$  \\
  $\Gamma_2$ ($T_d$) & $\Gamma_2$ & $\Gamma_1$ & $\Gamma_2$ \\
  $\Gamma_3$ ($T_d$) & $\Gamma_1+\Gamma_2$ & $\Gamma_2+\Gamma_3$ &
    $\Gamma_1+\Gamma_2$ \\
  $\Gamma_4$ ($T_d$) & $\Gamma_1+\Gamma_3+\Gamma_4$ & $\Gamma_1+
    \Gamma_2+\Gamma_3$ & $\Gamma_1+2\Gamma_2$ \\
  $\Gamma_5$ ($T_d$) & $\Gamma_2+\Gamma_3+\Gamma_4$ & $\Gamma_1+
    \Gamma_2+\Gamma_3$ & $2\Gamma_1+\Gamma_2$ \\
  \hline
  \hline
  \end{tabular}
\end{table}

To deduce the magnetic field induced splitting of the levels, we
employ first order degenerate perturbation theory. As mentioned
before, it is assumed that the separation of the levels is large
compared to the splitting caused by the field, so only the
subspace of Hilbert space connected to the level under
consideration needs to be taken into account. Given a set of basis
functions $|i\rangle$ for a particular level, we find the
corresponding sub-matrix $\langle i|\mathcal{H}_Z|j\rangle$ of
$\mathcal{H}_Z$ and diagonalize it to obtain the splitting as a
function of $\bB$.

Instead of trying to calculate matrix elements from
$\mathcal{H}_Z$ (after choosing a suitable set basis functions) it
is much more convenient to use the well-established approach of
constructing an \emph{effective Zeeman Hamiltonian}
\cite{tsukerblat94}. This comprises the construction of a matrix
of the required size, exploiting required symmetries to find
vanishing elements and relations between elements. The result is a
matrix that usually depends on a small number of unknown
phenomenological parameters, in terms of which the level splitting
can be expressed. This approach is especially advantageous in the
present situation, where both the values of the parameters
occurring in $\mathcal{H}_Z$ and the unperturbed wave functions
are not (exactly) known.

\section{Linear Zeeman effect}

In this section, we will investigate the first order Zeeman effect
of all the double acceptor levels mentioned before.

\subsection{The $\Gamma_i$-levels}

Because $\Gamma_4$ occurs in neither of the anti-symmetric direct
products $\{\Gamma_i\times\Gamma_i\}$ ($i=1\ldots 3$), the
effective Hamiltonian matrix $\mathcal{H}_{\mathrm{eff,lin}}$
vanishes identically for the three levels $\Gamma_i$. Hence, none
of these levels experiences a linear Zeeman effect.

The linear part of the effective Zeeman Hamiltonian for a
$\Gamma_4$ or $\Gamma_5$ level is given by
\cite{luttinger56,bhattacharjee72}
\begin{equation*}
    \mathcal{H}_{\mathrm{eff,lin}}=
      \mu_Bg(\mcB_xJ_x+\mcB_yJ_y+\mcB_zJ_z).
\end{equation*}
Here, $\mu_B$ is the Bohr-magneton and $J_x$, $J_y$ and $J_z$ are
matrix representations of the components of the angular momentum
operator with respect to some convenient basis. The components
$J_\alpha$ ($\alpha=x,y,z$) transform according to the $\Gamma_4$
representation of $\bar{T}_d$. Because
$\{\Gamma_4\times\Gamma_4\}=\{\Gamma_5\times\Gamma_5\}= \Gamma_4$,
the $\Gamma_4$ and $\Gamma_5$ level do have a linear Zeeman
effect. Calculating the eigenvalues of the matrix
$[\mathcal{H}_{\mathrm{eff,lin}}]_i$ ($i=4,5$) yields
$$
  \Delta E = \left\{\begin{array}{l}
                      +\mu_Bg\mcB \\
                      0 \\
                      -\mu_bg\mcB. \\
                    \end{array}\right.
$$
The eigenvalues are independent of the direction of the magnetic
field and hence give rise to an isotropic splitting.

\subsection{The $\Gamma_3+\Gamma_5$ level}

The situation where the zero-field splitting of the
$\Gamma_3+\Gamma_5$ level is small compared to the Zeeman energy
must be dealt with separately. Because
$\Gamma_3\times\Gamma_5=\Gamma_4+\Gamma_5$ contains $\Gamma_4$,
there are non vanishing cross-terms in the linear effective Zeeman
Hamiltonian for a $\Gamma_3\Gamma_5$-level. Therefore, such a
level will have a linear Zeeman shift different from that of the
individual $\Gamma_3$ and $\Gamma_5$ levels. The Hamiltonian
sub-matrix for the $\Gamma_3+\Gamma_5$ level is given by
\footnote{Because $\Gamma_3\times\Gamma_5 = \Gamma_4+\Gamma_5$ it
is possible (e.g. by using the coupling coefficients for
$\Gamma_3\times\Gamma_5$ as given in Ref.\onlinecite{koster63} and
inverting) to express all six possible products of $\Gamma_3$ and
$\Gamma_5$ wave functions as a linear combination of $\Gamma_4$
and $\Gamma_5$ wave functions. Using the Wigner-Eckart
orthogonality theorem and the fact that all operators occurring in
the linear Zeeman Hamiltonian transform according to the rows of
$\Gamma_4$, the matrix form can be derived.}
\begin{equation*}
  \left(\begin{array}{c|c}
    \emptyset & \begin{array}{ccc}
                  -\frac{\sqrt{3}}{2}a_{35}\mcB_x & \frac{\sqrt{3}}{2}a_{35}\mcB_y & 0 \\
                  -\frac{1}{2}a_{35}\mcB_x & -\frac{1}{2}a_{35}\mcB_y & a_{35}\mcB_z \\
                \end{array} \\
    \hline
    \begin{array}{cc}
      -\frac{\sqrt{3}}{2}a_{35}\mcB_x & -\frac{1}{2}a_{35}\mcB_x \\
      \frac{\sqrt{3}}{2}a_{35}\mcB_y  & -\frac{1}{2}a_{35}\mcB_y \\
      0                         & a_{35}\mcB_z \\
    \end{array} & [\mathcal{H}_\mathrm{eff,lin}]_5 \\
  \end{array}\right),
  \label{eq:mat35}
\end{equation*}
where $a_{35}$ is assumed to be real. This $5\times 5$-matrix is
given with respect to a basis with two $\Gamma_3$ wave functions
and three $\Gamma_5$-wave functions. Only the matrix-elements
connecting $\Gamma_3$ functions to $\Gamma_5$ functions are shown
explicitly. The upper left and bottom right parts are the matrices
of the individual $\Gamma_3$ and $\Gamma_5$ level, respectively,
as given in the previous subsection.

From this matrix, we determine the eigenvalues for $\bB$ parallel
to the main crystallographic directions. For $\bB\parallel \langle
100\rangle$, so $\mcB_x=\mcB$, $\mcB_y=\mcB_z=0$, we find
$$
  \Delta E=\left\{\begin{array}{l}
             0   \\
             \pm \mu_Bg\mcB\\
             \pm a_{35}\mcB.\\
           \end{array}\right.
$$
For $\bB\parallel \langle 111\rangle$, so
$\mcB_x=\mcB_y=\mcB_z=\mcB/\sqrt{3}$, we have
$$
  \Delta E=\left\{\begin{array}{l}
             0   \\
             \pm\frac{1}{2}\mu_Bg\mcB\pm\frac{1}{2}\mcB\sqrt{\mu_B^2g^2+2a_{35}^2}.\\
           \end{array}\right.
$$
Finally for $\bB\parallel \langle 110\rangle$, so
$\mcB_x=\mcB_y=\mcB/\sqrt{2}$ and $\mcB_z=0$, it is found that
$$
  \Delta E=\left\{\begin{array}{l}
             0   \\
             \pm\frac{1}{2}a_{35}\mcB \\
             \pm\frac{1}{2}\mcB\sqrt{2\mu_B^2g^2+3a_{35}^2}.\\
           \end{array}\right.
$$
We conclude that there is indeed a linear Zeeman effect in the
$\Gamma_3+\Gamma_5$ level and the size of the effect is dependent
of the direction of the field with respect to the crystal.

\section{Quadratic Zeeman effect}

For some of the levels we will also give a second order approach,
using the quadratic part of the effective Hamiltonian
$\mathcal{H}_{\mathrm{eff,quad}}$. Note that
$\mathcal{H}_{\mathrm{eff,quad}}$ contains both a second order
approach to the linear part of the original $\mathcal{H}_Z$ and a
first order approach to the quadratic part of the original
$\mathcal{H}_Z$.

\subsection{The $\Gamma_1$ and $\Gamma_2$ levels}

For the $\Gamma_1$ level, the effective quadratic Zeeman
Hamiltonian contains only one term and is straightforwardly given
by
$$
  \mathcal{H}_{\mathrm{eff,quad}}=a_1\mathcal{B}^2,
$$
where $a_1$ is a phenomenological parameter. The simple conclusion
is that a $\Gamma_1$ level will experience a quadratic shift,
independent of the direction of the magnetic field: $\Delta E =
a_1\mathcal{B}^2$. From this purely symmetry-based analysis,
conclusions can be drawn neither about the magnitude of $a_1$ nor
about its sign (that is, whether the state is diamagnetic or
paramagnetic). Because $\Gamma_2\times\Gamma_2=\Gamma_1$, a
similar expression holds for a $\Gamma_2$ level.

\subsection{The $\Gamma_3$ level}

For a $\Gamma_3$ level, the effective Hamiltonian contains two
unknown parameters and is given by \cite{kartheuser73,washimiya70}
\begin{equation*}
  \begin{split}
    \mathcal{H}_{\mathrm{eff,quad}}=a_3\mathcal{B}^2
    + b_3\big[-(2\mcB_z^2-&\mcB_x^2-\mcB_y^2)\sigma_x \\
    & + \sqrt{3}(\mcB_x^2-\mcB_y^2)\sigma_y\big], \\
  \end{split}
\end{equation*}
where $\sigma_x$ and $\sigma_y$ are Pauli spin matrices and $a_3$
and $b_3$ are phenomenological parameters. When
$\bB\parallel\langle 100\rangle$ the eigenvalues are given by
$$
  \Delta E=(a_3\pm 2b_3)\mcB^2.
$$
This is a symmetric quadratic splitting superimposed on a
quadratic shift. When $\bB\parallel\langle 111\rangle$, there is
only one eigenvalue
$$
  \Delta E=a_3\mcB^2,
$$
meaning that there is no splitting in second order and the
quadratic shift is the same as for $\bB\parallel\langle
100\rangle$. Finally, for $\bB\parallel\langle 110\rangle$, we
find the eigenvalues
$$
  \Delta E=(a_3\pm b_3)\mcB^2.
$$
The Zeeman effect for this field direction is similar to
$\bB\parallel\langle 100\rangle$, but the splitting is twice as
small.

\subsection{The $\Gamma_4$ and $\Gamma_5$ levels}

Because the symmetrized squares of $\Gamma_4$ and $\Gamma_5$
satisfy $[\Gamma_4\times\Gamma_4]=[\Gamma_5\times\Gamma_5]$, the
results for the $\Gamma_4$ and $\Gamma_5$ levels are similar. For
these two levels, the quadratic part of the effective Zeeman
Hamiltonian has three unknown parameters $a_i$, $b_i$ and $c_i$
($i=4,5$) and is given by \cite{bhattacharjee72}
\begin{equation*}
  \begin{array}{l}
    \mathcal{H}_{\mathrm{eff,quad}}=a_i\mathcal{B}^2\\
      \ \ \ +b_i\left[3(\mcB_x^2J_x^2+\mcB_y^2J_y^2-\mcB_z^2J_z^2)-2\mcB^2\right]\\
      \ \ \ +c_i\big[\mcB_y\mcB_z\{J_y,J_z\}+
               \mcB_x\mcB_z\{J_x,J_z\} \\
      \ \ \ \ \ + \mcB_x\mcB_y\{J_x,J_y\}\big],\\
  \end{array}
\end{equation*}
where $\{A,B\}=\frac{1}{2}(AB+BA)$ denotes the anti-commutator of
$A$ and $B$. We calculate the eigenvalues of the full quadratic
Hamiltonian matrix $[\mathcal{H}_{\mathrm{eff,lin}}]_i +
[\mathcal{H}_{\mathrm{eff,quad}}]_i$ for the three main
crystallographic directions. For $\bB\parallel\langle 100\rangle$
we have
$$
  \Delta E=\left\{\begin{array}{l}
             \mu_Bg\mcB+(a_i+b_i)\mcB^2 \\
             (a_i-2b_i)\mcB^2 \\
             -\mu_Bg\mcB+(a_i+b_i)\mcB^2 \\
           \end{array}\right.
$$
For $\bB\parallel\langle 111\rangle$ we have
$$
  \Delta E=\left\{\begin{array}{l}
             \mu_Bg\mcB+(a_i+\frac{1}{6}c_i)\mcB^2\\
             (a_i-\frac{1}{3}c_i)\mcB^2\\
             -\mu_Bg\mcB+(a_i+\frac{1}{6}c_i)\mcB^2.\\
           \end{array}\right.
$$
And for $\bB\parallel\langle 110\rangle$ we find
$$
  \Delta E=\left\{\begin{array}{l}
             \mcB\sqrt{\mu_B^2g^2+(\frac{1}{8}c_i+\frac{3}{4}b_i)\mcB^2}\\
             \ \ \ \ \ +(a_i+\frac{1}{4}b_i+\frac{1}{8}c_i)\mcB^2\\[5pt]
             (a_i-\frac{1}{2}b_i-\frac{1}{4}c_i)\mcB^2\\[5pt]
             -\mcB\sqrt{\mu_B^2g^2+(\frac{1}{8}c_i+\frac{3}{4}b_i)^2\mcB^2}\\
             \ \ \ \ \ +(a_i+\frac{1}{4}b_i+\frac{1}{8}c_i)\mcB^2\\
           \end{array}\right.
$$
It follows that in second order the spitting is no longer
symmetric and isotropic for these levels.

\begin{figure}
  \centering
  \includegraphics[width=8.6cm]{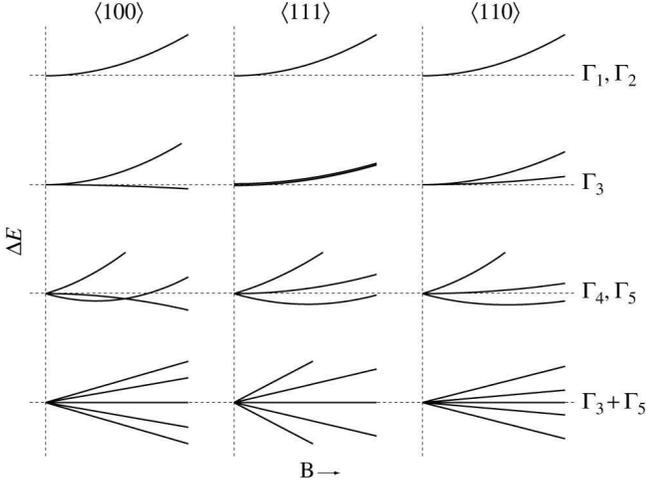}
  \caption{Schematic overview of level splitting in a magnetic field
  along the main crystallographic axes in several kinds of two-hole
  acceptor levels.
  The figure illustrates the qualitative
  aspects of the splitting. The values of the parameter have been chosen
  to emphasize these features.}
  \label{fig:Bcalc}
\end{figure}

The Zeeman effect of the levels treated so far is schematically
illustrated in Fig.~\ref{fig:Bcalc}.

\section{Central field approximation}

Finally we discuss the Zeeman effect for two-hole states in the
central field approximation. In this approximation, we must start
from the one-hole levels and their behavior in a magnetic field.
The two-hole wave functions are anti-symmetrized products of
one-hole wave functions and the energy levels are obtained by
examining the various ways to put the two holes in the one-hole
levels.

We will present results for the case where both holes are put in a
$\Gamma_8$ level and $\bB\parallel\langle 100\rangle$ only.
Similar results for the other types of levels and other directions
of the field are easily obtained in an analogous way.

For a magnetic field $\bB\parallel \langle 100\rangle$, the single
hole $\Gamma_8$ ground state is split into $\Gamma_5$, $\Gamma_6$,
$\Gamma_7$ and $\Gamma_8$-levels of $\bar{S}_4$
\cite{bhattacharjee72}. Because holes are fermions, each of these
non-degenerate levels can be occupied by at most one hole. By
putting each of the two holes in a different level, this gives
rise to six two-hole levels $\Gamma_5 \times \Gamma_6 = \Gamma_1$,
$\Gamma_5 \times \Gamma_7 = \Gamma_3$, $\Gamma_5 \times \Gamma_8 =
\Gamma_2$, $\Gamma_6 \times \Gamma_7 = \Gamma_2$, $\Gamma_6 \times
\Gamma_8 = \Gamma_4$, and $\Gamma_7 \times \Gamma_8 = \Gamma_1$,
where all representations are of $\bar{S}_4$.

The energy shifts of the single-hole levels have been determined
experimentally \cite{bhattacharjee72}. The shifts of the two-hole
levels can be calculated as the sum of the shifts of the
individual single hole levels from which they are composed. This
results in a linear shift for each two-hole level, given by $\mu_B
g \mcB$, with $g=\frac{3}{2}g_{3/2} + \frac{1}{2}g_{1/2}$ for
$\Gamma_2$, $g=\frac{3}{2}g_{3/2} - \frac{1}{2}g_{1/2}$ for
$\Gamma_4$, $g=0$ for $2\Gamma_1$, $g=-\frac{3}{2}g_{3/2} +
\frac{1}{2}g_{1/2}$ for $\Gamma_3$ and $g=-\frac{3}{2}g_{3/2} -
\frac{1}{2}g_{1/2}$ for $\Gamma_2$. The parameters $g_{3/2}$ and
$g_{1/2}$ are the $g$-factors for the single hole $j=\frac{3}{2}$
and $j=\frac{1}{2}$-levels respectively. Experimental values for
B$^0$ in Si are $g_{3/2}=1.12$ and $g_{1/2}=1.04$ \cite{merlet75}.
In the above, a small overall shift is neglected.

\section{Application to B$^+$}

Only states arising from $\{\Gamma_8\times\Gamma_8\}$ (see
Table~\ref{tab:twohole}) are candidates for the B$^+$ ground
state. These are $\Gamma_1$, $\Gamma_3$, $\Gamma_5$,
$\Gamma_3+\Gamma_5$ and the unsplit (central field)
$\{\Gamma_8\times\Gamma_8\}$. Each of these five possibilities for
the B$^+$ ground state will be compared to existing experimental
data. From the previous section, we conclude that all possible
ground state levels behave qualitatively differently in a magnetic
field. Therefore, it is in principle possible to determine the
nature of the actual ground state of B$^+$ from the analysis of a
sufficiently detailed experiment. Though this approach is hampered
by the fact that the value of the parameters are not known, it is
possible to draw conclusions based on the qualitative
characteristics, such as linear or quadratic splitting/shift and
the asymmetry of the splitting.

\begin{figure}
  \centering
  \includegraphics[width=8.6cm]{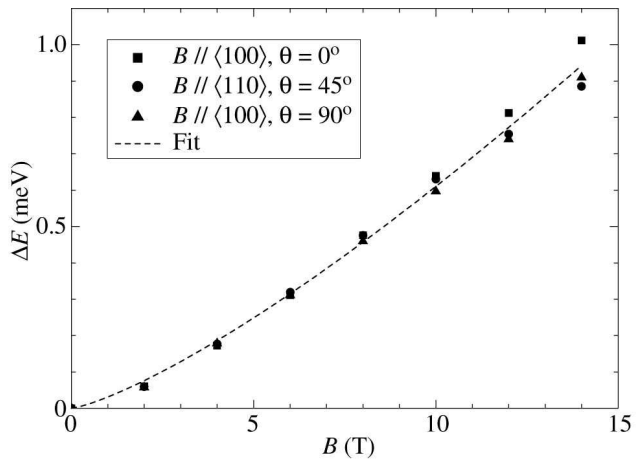}
  \caption{Magnetic field induced shift of the B$^+$ ground state
  deduced from electrical transport measurements \cite{caro03}. The angle
  between the magnetic field and the current is denoted by $\theta$.
  The expression $\Delta E\propto \mcB(1+\alpha\mcB)$ was fit to the
  data, yielding $\alpha=0.046~\mathrm{T}^{-1}$.}
  \label{fig:Bmeas}
\end{figure}

We refer to our recent experiments reported in
Ref.~\onlinecite{caro03} and summarize the main observations. The
ground state energy shifts upwards (that is, in the direction of
the valence band) and is therefore diamagnetic. The shift has both
a linear and a quadratic component. The total shift amounts to
1~meV at a magnetic field of 14~T and was equal for the $\langle
100\rangle$ and $\langle 110\rangle$ directions (see
Fig.~\ref{fig:Bmeas}). The width of the observed peak (full width
at half maximum) increased from 1.2~meV to 1.5~meV in the same
magnetic field range. Within the experimental error ($\sim
0.2$~meV), no splitting of the peak was detected \footnote{In
these experiments the concentration of B-impurities in the silicon
was so high that they cannot be considered as fully isolated as
proven by the increased binding energy of the second hole in the
B$^+$-state. The interaction of a B$^+$ state with neighboring
B$^0$ states is however not expected to change the nature of its
ground state.}.

The experimentally observed super linear overall shift,
independent from the direction of $\bB$, best matches the behavior
of a $\Gamma_1$ state, although this leaves the strong linear
component in the measured magnetic field dependence unexplained.
Therefore, we believe that the ground state of B$^+$ is indeed a
$\Gamma_1$ state. This hypothesis does imply that the observed
linear component in the peak shift and the peak broadening are is
due to other processes (e.g.\ the Stark effect) as already
suggested in Ref.~\onlinecite{caro03}.

The broadening in the observed peak is linear in the magnetic
field and independent of its direction. Therefore, it cannot be
explained as unresolved splitting of a $\Gamma_3$ level. A
$\Gamma_5$ or $\Gamma_3+\Gamma_5$ ground state would give rise to
linear splitting (broadening), but no overall shift would be
expected  in first order. Moreover, the magnitude of the splitting
in a $\Gamma_3+\Gamma_5$ level would depend on the magnetic field
direction. Therefore, these possibilities are not consistent with
the experimental observations. Only when the parameter $a_3$
($a_5$) would much larger than all other relevant parameters (that
is $a_3\ll b_3$ or $a_5\ll b_5,c_5,\mu_bg/\mcB$), the magnetic
field dependence of the $\Gamma_3$ ($\Gamma_5$) state would be
similar to that of the $\Gamma_1$ state. In that case, $\Gamma_3$,
$\Gamma_5$ and $\Gamma_3+\Gamma_5$ states cannot be rejected as
potential ground state symmetries for B$^+$.

The central field approach is unlikely to yield good results for
B$^+$, for which the wave functions of the two holes are expected
to overlap considerably (due to the small nuclear charge). The
peak splitting (or broadening, due to unresolved splitting)
expected in this approach between the two $\Gamma_2$ levels would
be given by $2\mu_B(\frac{3}{2}g_{3/2} + \frac{1}{2}g_{1/2})B$.
Assuming the B$^0$-values of the $g$-factors are valid here, this
would amount to 3.6~meV for $B=14$~T. This is much larger than the
observed 0.3~meV increase of the FWHM of the measured resonance
peak. Moreover, the 1~meV shift observed in the experiment is much
larger than the expected overall peak shift in this approach.
Therefore, the description of the B$^+$ ground state in the
central field approximation is not consistent with the
experimental observations.

In summary, magnetic field dependent measurements indicate that
the B$^+$ ground state is a non-degenerate $\Gamma_1$ state. It
would be interesting to have higher resolution spectroscopy data
available, in order to exclude that the observed peak broadening
is due to unresolved splitting. Knowledge of the B$^+$ ground
state wave function would allow for obtaining quantitative
information about the phenomenological parameters, which would be
advantageous in the interpretation of experimental data.

\section{Conclusions}

In conclusion, we have presented a general group theoretical study
of the magnetic field dependence of two-hole states of acceptors
in tetrahedral semiconductors. We have used our results to analyze
recent experimental observations. This analysis indicates that the
B$^+$ ground state is most compatible with a $\Gamma_1$ state.

\begin{acknowledgments}
We thank I.~D.~Vink for useful discussions. One of us, S.R.,
wishes to acknowledge the Royal Netherlands Academy of Arts and
Sciences for financial support. This work is part of the research
program of the Stichting voor Fundamenteel Onderzoek der Materie,
which is financially supported by the Nederlandse Organisatie voor
Wetenschappelijk Onderzoek.
\end{acknowledgments}

\end{document}